# Quantum oscillations of the critical current and high-field superconducting proximity in ballistic graphene

M. Ben Shalom[1], M. J. Zhu[1], V. I. Fal'ko[2], A. Mishchenko[1], A. V. Kretinin[1], K. S. Novoselov[1], C. R. Woods[1], K. Watanabe[3], T. Taniguchi[3], A. K. Geim[1], J. R. Prance[2]

[1]School of Physics & Astronomy, University of Manchester, Oxford Road, M13 9PL Manchester, UK
[2]Department of Physics, University of Lancaster, Lancaster, UK
[3]National Institute for Materials Science, 1-1 Namiki, Tsukuba, 305-0044 Japan

*Graphene-based Josephson junctions provide a novel platform for studying the proximity effect[1-3] due to graphene's unique electronic spectrum and the possibility to tune junction properties by gate voltage[3-16]. Here we describe graphene junctions with a mean free path of several micrometers, low contact resistance and large supercurrents. Such devices exhibit pronounced Fabry-Pérot oscillations not only in the normal-state resistance but also in the critical current. The proximity effect is mostly suppressed in magnetic fields below 10 mT, showing the conventional Fraunhofer pattern. Unexpectedly, some proximity survives even in fields higher than 1 T. Superconducting states randomly appear and disappear as a function of field and carrier concentration, and each of them exhibits a supercurrent carrying capacity close to the universal quantum limit[17,18]. We attribute the high-field Josephson effect to mesoscopic Andreev states that persist near graphene edges. Our work reveals new proximity regimes that can be controlled by quantum confinement and cyclotron motion.*

The superconducting proximity effect relies on penetration of Cooper pairs from a superconductor (S) into a normal metal (N) and is most pronounced in systems with transparent SN interfaces and weak scattering so that superconducting correlations penetrate deep inside the normal metal. Despite being one atom thick and having a low density of states, which vanishes at the Dirac point, graphene (G) can exhibit low contact resistance and ballistic transport on a micrometer scale[19-21], exceeding a distance between superconducting leads by an order of magnitude. These properties combined with the possibility to electrostatically control the carrier density $n$ offer tunable Josephson junctions in a regime that can be referred to as ballistic proximity superconductivity[22]. Despite recent intense interest in SGS devices[3-16] that can show features qualitatively different from the conventional SNS behavior[1], the ballistic proximity regime remains little studied.

Our SGS devices are schematically shown in Fig. 1 and described in further detail in Supplementary Section 1. The essential technological difference from the previously studied SGS junctions[3-15] is the use of graphene encapsulated between boron-nitride crystals[20,21] and a new nanostrip geometry of the contacts. This allows a high carrier mobility, low charge inhomogeneity and low contact resistance. More than twenty SGS junctions with the width $W$ between 0.5 and 8 µm and the length $L$ between 0.15 and 2.5 µm were studied, showing reproducible behavior and consistent changes with $L$ and $W$. First, we characterize the devices above the transition temperature $T_C \approx 7$ K of our superconducting contacts. Figure 1b shows examples of the normal-state resistance $R_n$ as a function of back gate voltage $V_g$ that changes $n$ in graphene. The neutrality point (NP) was found shifted to negative $V_g$ by a few V, with the shift being consistently larger for shorter devices. This is due to electron doping induced by our contacts. For ballistic graphene, such doping is uniform over the entire device length[23], and the observed smearing of such $R_n(V_g)$ curves near the NP allows an estimate for charge inhomogeneity as $\approx 2\times 10^{10}$ cm$^{-2}$. For consistency, data for devices with different $L$ are presented as a function of $\Delta V_g$, the gate voltage counted from the NP.

For positive $\Delta V_g$ (electron doping) and $n > 10^{11}$ cm$^{-2}$, devices with the same $W$ exhibit the same $R_n(\Delta V_g)$ dependence, independently of $L$ (Fig. 1b). This shows that the mean free path in graphene is limited by the contact separation and yields charge carrier mobility >300,000 cm$^2$ V$^{-1}$ s$^{-1}$, in agreement with the quality measured for similarly made Hall bar devices. The dashed curve in Fig. 1b indicates the behavior expected in the quantum ballistic limit, $R_Q = (h/e^2)/4N$ where $N = \text{int}(2W/\lambda_F)$ is the number of propagating electron modes, $\lambda_F$ the Fermi wavelength and the factor 4 corresponds to graphene's degeneracy. The difference between $R_Q$ and the experimental curves yields a record low contact resistivity of ≈35 Ohm μm. This value corresponds to an angle-averaged transmission probability $Tr \approx 0.8$ (Supplementary Section 2).

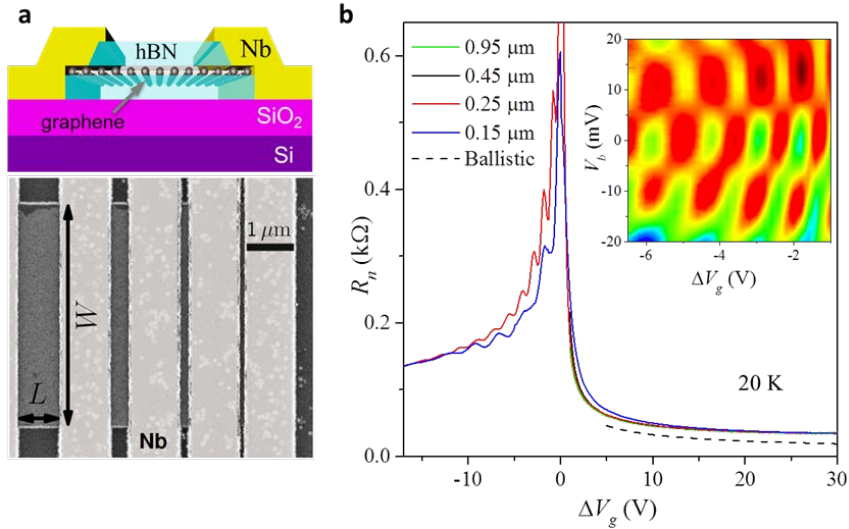

**Figure 1| Ballistic SGS junctions. a**, Top: Junctions' schematics. Bottom: Electron micrograph of a set of such junctions with different $L$. A few nm-wide graphene ledge (top drawing) is referred to as a nanostrip contact. **b**, Typical behavior for SGS junctions with different $L$ but same $W$ =5 μm. To avoid an obscuring overlap between four oscillating curves at negative $\Delta V_g$, we plot $R_n$ only for the two shortest junctions. For electron doping higher than ~10$^{11}$ cm$^{-2}$, the four curves overlap within the line width. The dashed curve shows $R_Q$. Inset: Changes in the differential conductance $dI/dV$. $L$ =0.25 μm. Color scale: -1 to 1 mS.

For hole doping, $R_n$ becomes significantly higher indicating smaller $Tr$ because of pn junctions that appear at the Nb contacts and lead to partial reflection of electron waves. This effectively creates a Fabry-Pérot (FP) cavity[24,25]. The standing waves lead to pronounced oscillations in $R_n$ as a function of both $V_g$ and applied bias $V_b$ (see Fig. 1b). The oscillatory behavior indicates that charge carriers can cross the graphene strip several times preserving their monochromaticity and coherence. Some FP oscillations could also be discerned for positive $\Delta V_g$ but they were much weaker because of higher $Tr$. The observed FP behavior in the normal state agrees with the earlier reports[24,25]. Its details can be modelled accurately if we take into account that the position of pn junctions varies with $V_g$ so that the effective length of the FP interferometer becomes notably shorter than $L$ at low hole doping (Supplementary Section 3).

After characterizing our SGS devices at temperatures $T$ above $T_C$, we turn to their superconducting behavior. All of them, independently of $L$, exhibited the fully developed proximity effect as shown in Figs. 2a,b. The critical current $I_c$ remained finite at the NP and rapidly increased with $|\Delta V_g|$, reaching densities >5 μA/μm for high electron doping, notably larger than previously reported[3-16]. Such high $I_c$ are due to ballistic transport and low contact resistance. Indeed, $I_c$ can theoretically reach a value[1,26]

$$I_c = \alpha\Delta/eR_n \qquad (1)$$

with $\alpha \approx 2.07$. Because in our devices $R_n \approx R_Q = h/4Ne^2$ the equation implies that we approach the quantum limit $I_c \approx (e\Delta/h)4N$ where the supercurrent is determined solely by the number of propagating Cooper-pair modes each having the supercurrent carrying capacity[2,17],

$$I_Q \approx e\Delta/h \qquad (2).$$

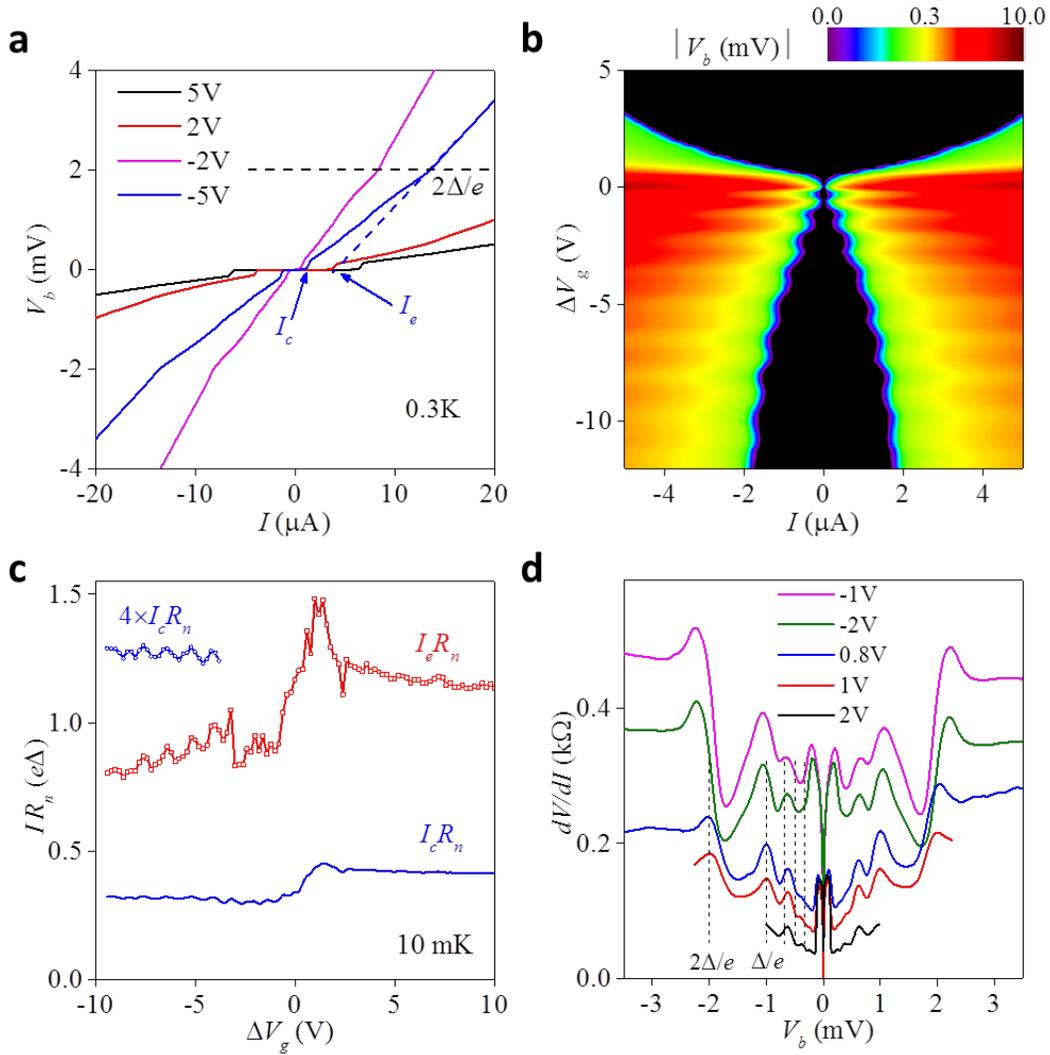

**Figure 2| Quantum oscillations in supercurrent. a**, Examples of I-V characteristics for ballistic SGS junctions in the superconducting state. The data are for the device in Fig. 1 with $L$ =0.25 µm. The arrows explain notions $I_c$ and $I_e$. **b**, Absolute voltage $|V_b|$ across the SGS junction in (a) for a wide range of doping. The black region corresponds to the zero-resistance state, and its edge exhibits FP oscillations. **c**, $I_cR_n$ and $I_eR_n$ for a device with $L$ =0.3 µm; $W$ =6.5 µm. Each data point is extracted from a trace such as in (a). Upper inset: Oscillatory part of $I_cR_n$ is magnified. Similar behavior was observed for other devices including the one in (b). **d**, Differential resistance calculated from I-V curves such as in (a). The expected peak positions due to multiple Andreev reflection are shown by dashed lines, using $\Delta$ found from $T_C$.

Eq. (1) suggests that $I_cR_n$ should be a constant. This holds well in our SGS devices away from the NP (Fig. 2c) and indicates that external noise, fluctuations and other mechanisms[3-16] whcih are dependent on $n$ or $R_n$ do not limit $I_c$. However, Fig. 2c shows $\alpha \approx 0.3$-$0.4$, roughly six times smaller than $\alpha$ in eq. (1). One of the reasons is multiple Andreev reflections[1,27,28] that lead to an earlier onset of dissipation (Fig. 2d) and are not accounted for[26] in deriving eq. (1). In the absence of multiple Andreev processes, $I_c$ should be close to the excess current $I_e$, which is determined as shown in Fig. 2a. Figure 2c presents $I_eR_n$ which yields $\alpha \approx 1$. The remaining disagreement is attributed to the fact that our devices are not in the limit of short $L$. This is evidenced from the fact that $I_cR_n$ continues to grow with decreasing $L$ down to 0.15 µm, our shortest junctions (Supplementary Section 4). The $L$ dependence reflects decoherence of Cooper pairs and can be associated with an energy scale $E_{T*} = \hbar v_F/\Lambda$, analogous to the Thouless energy in diffusive junctions[1,26]. The distance $\Lambda$ that Cooper pairs cover during their transport through graphene can be estimated as $L/Tr \sim 1$µm, yielding $E_{T*} \sim 1$ meV comparable with $\Delta$. In this intermediate regime, smaller $I_eR_n$ are expected[26]. This explanation is also consistent with somewhat larger $I_{c,e}R_n$ observed for electron doping where $Tr$ and, hence, $E_{T*}$ are larger (Fig. 2c).

As a consequence of quantized transport in the normal state (Fig. 1), the supercurrent also exhibits quantum oscillations that are seen in Fig. 2b for negative $\Delta V_g$. Note that Eq. (1) suggests that such FP oscillations should occur simply because $R_n$ oscillates. Indeed, $R_n$ and $I_c$ are found to oscillate in antiphase, compensating each other in the final products $I_{c,e}R_n$. However, oscillations in the critical current are roughly 3 times larger in amplitude than those in $R_n$. Therefore, the quantization appears in both $I_cR_n$ and $I_eR_n$ (Fig. 2c) and is not a trivial consequence of oscillating $R_n$. Also, the observed oscillations cannot be explained by changes in $\Delta$ (Supplementary Section 4). We attribute the effect to changes in $E_{T*}$ which are caused by oscillating transparency of our FP resonators. Indeed, Cooper pairs are expected to spend less time in graphene at transmission resonances, that is, when $R_n$ exhibits minima. This leads to an increase in $E_{T*}$ and effectively shifts the SGS conditions closer to the short-$L$ limit of eq. (1). This explanation agrees with the observed phase of the oscillations in Fig. 2c.

In magnetic field $B$, our ballistic junctions exhibit further striking departures from the conventional behavior (Fig. 3). In small $B$ such that a few flux quanta $\phi_0 = h/2e$ enter an SGS junction, we observe the standard Fraunhofer dependence[1]

$$I_c = I_c(B=0) \, |\sin(\pi\Phi/\phi_0)/\Phi| \qquad (3)$$

where $\Phi = S \times B$ is the flux through the junction area $S = L \times W$. Marked deviations from eq. (3) occur in $B > 5$ mT (Fig. 3a). Figures 3b-e show that, in this regime, the supercurrent no longer follows the oscillatory Fraunhofer pattern but pockets of proximity superconductivity can randomly appear as a function of $n$ and $B$. At low doping, the pockets can be separated by extended regions of the normal state where no supercurrent could be detected with accuracy of a few nA << $I_Q$ (Figs. 3c,e). Within each pocket, I-V characteristics exhibit a gapped behavior (inset of Fig. 3d) with $I_c \approx I_Q \approx 40$ nA, although the exact value depends on doping and $I_c$ falls down to $\approx 10$ nA close to the NP, possibly due to a rising contribution of electrical noise that suppresses apparent $I_c$ (Fig. 3c). These proximity states persist until $B$ as large as $\approx 1$ T ($\Phi/\phi_0 \sim 10^3$) and are highly reproducible, although occasional flux jumps in Nb contacts can reset the proximity pattern (Supplementary Section 5). Correlation analysis presented in Supplementary Section 6 yields that, to suppress such superconducting states, it requires changes in $\Phi$ of $\approx \phi_0$ and changes in the Fermi energy of $\approx 1$ meV.

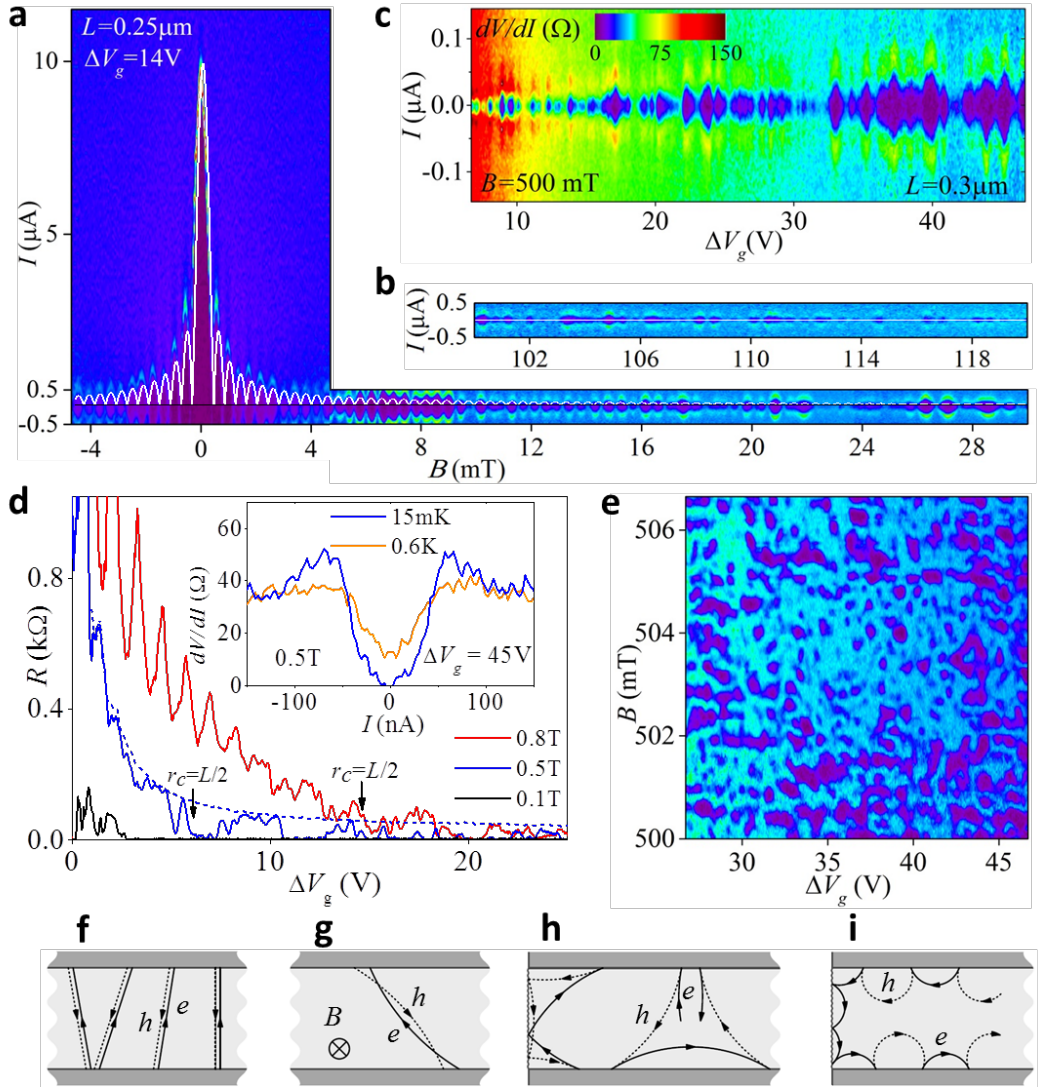

**Figure 3| Random proximity states. a** Example of $dV/dI$ as a function of applied current $I$ and $B$. The purple regions correspond to the zero-$R$ state and their edges mark the critical current $I_c$. The map is symmetric in both $I$ and $B$. The white curve is given by eq. (3). The low-$B$ periodicity is ≈0.4 mT, smaller than expected from the device's area, which is attributed to the Meissner screening that focuses the field into the junction[30]. **b**, Continuation of the map from (a) above 0.1 T. Intervals with finite $I_c$ continue randomly appear, despite the Fraunhofer curve is indistinguishable from zero. **c**, Another high-$B$ example but as a function of $\Delta V_g$ in 0.5 T. **d**, Examples of low-bias resistance ($I$ =2 nA) in different $B$. The dashed curve for 0.5 T shows that $I$ =150 nA > $I_Q$ completely suppresses superconductivity. The arrows mark the expected onset of edge state transport. **e**, Local map of fluctuating superconductivity. $T \approx$10 mK; all color scales are as in (c). Inset in (d): Typical I-V characteristics for high-$B$ superconducting states. **f-i**, Andreev states in ballistic junctions in zero (f), intermediate (g,h) and quantized $B$ (i). The states are suppressed by cyclotron motion in the bulk (g,h) but persist near edges (h).

To explain the fluctuating superconductivity pattern at large $B$, we take the view[1,26-28] that the proximity is mediated by Andreev bound states that consist of electrons and holes travelling along same trajectories but in opposite directions (Fig. 3f). In low $B$, interference between many Andreev states results in the oscillatory suppression of $I_c$ as described by eq. (3) and seen in Fig. 3a. Although not observed before, the Fraunhofer pattern in ballistic devices can be expected to break down because cyclotron motion deflects electrons and holes in opposite directions so that they can no longer retrace each other (Fig. 3g). We estimate the field required to suppress Andreev states in the bulk as $B^* \sim \Delta/eLv_F$ (Supplementary Section 7). For the devices in Fig. 3, this yields $B^* \approx 5$ mT, in agreement with the field where strong deviations from eq. (3) are observed.

The random pockets of superconductivity found for $B \gg B^*$ are unexpected because cyclotron motion strictly forbids Andreev bound states, beyond the suppression mechanism described by eq. (3). Although Andreev states cannot exist in the ballistic bulk at high $B$, the transfer of Cooper pairs is still possible near graphene edges that diffusively scatter charge carriers (Supplementary Section 7). An example of trajectories responsible for such near-edge Andreev states is depicted in Fig. 3h. These states are a natural candidate for explaining the observed Josephson effect with $I_c \sim e\Delta/h$, characteristic of a single Andreev bound state[17]. Such states are expected to be suppressed by changes in $\Phi$ of $\approx \phi_0$, the scale at which interference patterns alter in phase-coherent devices[18,29,30]. Similarly[18], a favorable interference pattern should change with changing the Fermi energy by $E_T^* \approx 1$ meV. Both scales are in good agreement with the experiment.

The near-edge proximity effect is expected to be terminated only if the cyclotron radius $r_c$ becomes shorter than $L/2$. This condition is marked in Fig. 3d and seen more clearly in the data of Supplementary Section 8. Finally, we note that that the edge superconductivity was not observed for hole doping, which can be attributed to the fact that Klein tunneling in graphene collimates trajectories perpendicular to the pn interface[24], making it essentially impossible to form closed-loop Andreev states shown in Fig. 3h (Supplementary Section 7). In principle, the effect of near-edge Andreev states could be further enhanced by presence of extended electronic states at graphene edges but no evidence for the latter could be found experimentally (Supplementary Section 9).

**Methods**

The measurements were carried out in a helium-3 cryostat for $T$ down to 0.3 K and in a dilution refrigerator, for lower $T$. All electrical connections to the sample passed through cold RC filters (Aivon Therma) and additional ac filters were on the top of the cryostats. The differential resistance was measured in the quasi-two-terminal geometry (using 4 superconducting leads to an SGS junction) and in the current-driven configuration using an Aivon preamplifier and a lock-in amplifier. To probe the superconducting proximity, we used an excitation current of 2 nA.

## Supplementary Sections

### 1. Device fabrication

Monolayer graphene was encapsulated between two relatively thick (typically, >30 nm) crystals of hexagonal boron nitride (hBN) by using the dry-peel transfer technique as detailed previously[S1]. The hBN-graphene-hBN stack was assembled on an oxidized Si wafer (300 nm of $SiO_2$) and then annealed at 300 °C in a forming gas (Ar-$H_2$ mixture) for 3 hours. As the next step we used the standard electron-beam lithography to make a PMMA mask that would define contact regions. Reactive ion etching (Oxford Plasma Lab 100) was employed to make trenches in the heterostructure through the mask. The etching process was optimized to achieve high etching rates for hBN with respect to both PMMA and graphene. We used a mixture of $CHF_3$ and $O_2$ which allowed rates of 300, 60 and 3 nm per min for hBN, PMMA and graphene, respectively. Importantly, the PMMA mask was not cross-linked during the etching and allowed easy liftoff so that metal contacts could be deposited directly after plasma etching. This procedure allowed us to avoid additional processing and, accordingly, contamination of the exposed graphene edges. The same etching recipe was later used to define the device geometry, which in this case was a constant width $W$ of our Josephson junctions.

Due to the large difference in the etching rates of graphene and hBN, the resulting edge profile was found to exhibit a step of, typically, 5 nm in width as depicted schematically in Fig. 1a of the main text and shown in micrographs of Fig. S1. This step developed because graphene effectively served as a mask during etching of the bottom hBN, leading to a gradual exposure of graphene buried under the top hBN. In comparison with contacts prepared in the same manner but without the highly selective etching, the graphene nanostrip provided a notably lower contact resistance (see below).

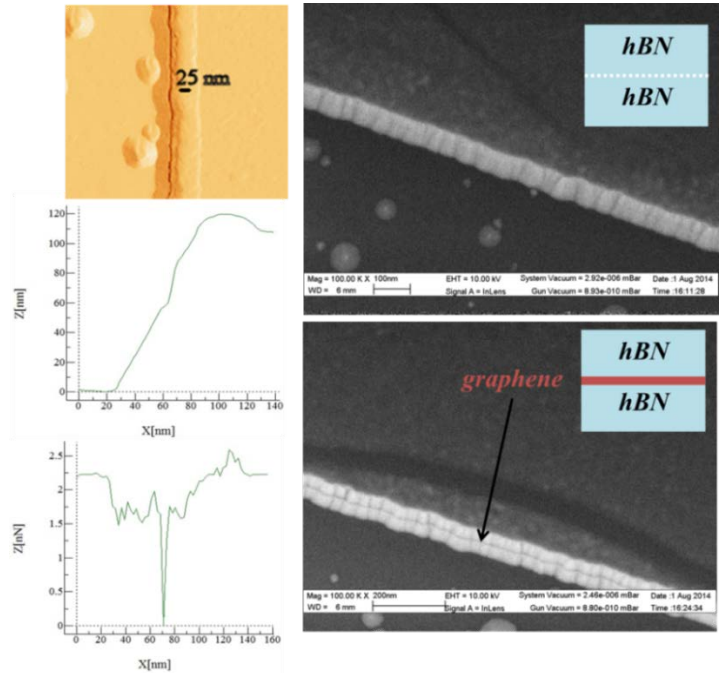

**Figure S1 | Nanostrip contacts to encapsulated graphene.** Left column: Atomic force microscopy of a plasma-etched edge of an hBN-graphene-hBN stack. Top: Imaging in the adhesion mode (color scale: 0 to 2.5 nN). The middle and bottom plots are typical topography and adhesion line scans across the step, respectively. A narrow graphene ledge appeared in the middle due to the large difference in etching rates between graphene and hBN. The right column shows scanning electron microscopy images of etched edges for hBN-hBN and hBN-graphene-hBN stacks. The narrow graphene step is clearly visible in the bottom image.

## 2. Superconducting contacts to graphene

As superconducting contacts, we used 50 nm thick films of Nb with an adhesion sublayer of Ta (5nm). Also, a few nm of Ta were put on top to protect Nb from oxidation. The trilayer film was deposited by radio-frequency sputtering at a rate of 5 nm/min and a base pressure of $\approx 10^{-9}$ Torr. The resulting films exhibited a sharp superconducting transition as shown in Fig. S2a. Here $T_C$ =7.2 K and the second critical field $H_{C2} \approx 3.5$ T, which yields the superconducting gap $\Delta$ =1.76$T_C \approx$12K and the coherence length $\xi = (\phi_0/2\pi H_{C2})^{1/2} \approx 10$ nm. The data are for the same set of SGS devices as in Fig. 1 of the main text. Variations in superconducting characteristics between different sets of SGS junctions did not exceed 10%.

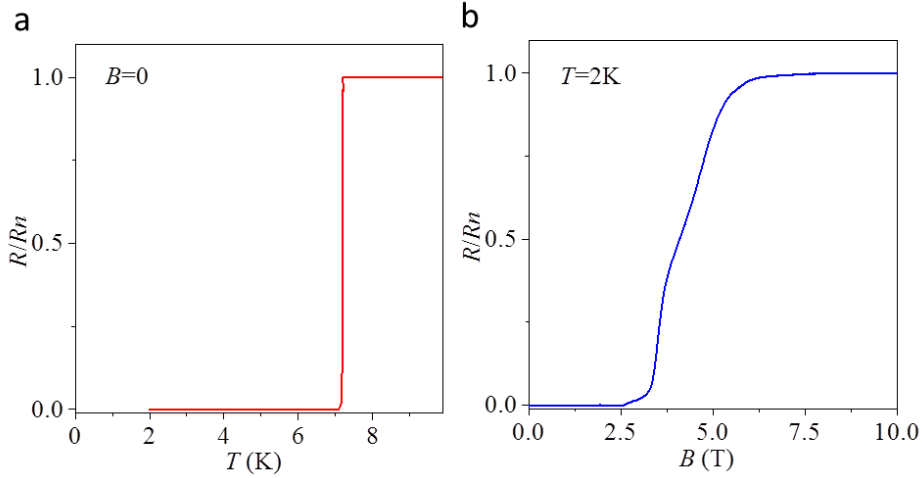

**Figure S2 | Characterization of Nb contacts.** Resistive measurements of their critical temperature (**a**) and critical field (**b**). $Rn$ is the normal-state resistance of our Nb/Ta films above $T_C$.

For a ballistic device with superconducting (zero-resistance) leads, the measured resistance is given by $R_n = R_Q + 2R_C$ where $R_Q$ is the quantum ballistic resistance determined in the main text and $R_C$ is the contact resistance per interface. Figure 1b of the main text plots $R_Q$ using the Fermi wavelength $\lambda_F$ that was calculated from carrier density $n$ induced by gate voltage. One can see that, independently of $L$, all the devices exhibited the same shift in $R_n$ upwards with respect to $R_Q$, which indicates a constant contribution, $2R_C$. For electron doping of graphene with $\Delta V_g \sim 10$ V, which corresponds to $n \approx 5\times 10^{11}$ cm$^{-2}$, $R_Q \approx 32$ Ohm for devices with $W$ =5 μm whereas we measured $R \approx 46$ Ohm. This yields $R_C \approx 7$ Ohm and contact resistivity of 35 Ohm μm. We find the same $R_C$ for all $\Delta V_g$ >10 V. The quality of our graphene-superconductor interface can also be characterized by their average transmission probability $Tr$ given by[S2] $Tr = R_Q/(R_Q + R_C)$. For $\Delta V_g \approx 10$ V, we calculate $Tr \approx 0.82$, that is, a highly transparent GS interface.

The low contact resistivity and high transmission probability of our nanostrip contacts were found to be highly reproducible for different devices, even though the etching and metal deposition required transfer between different equipment and, consequently, exposure of the interfaces to air. The nanostrip contacts' quality can be attributed to the finite width of the graphene-metal contact area (compared to so-called one-dimensional contacts[21]) and an extensive damage of the exposed graphene by oxygen plasma, which is known to improve contact quality[S3]. A good match between the work functions of Ta and damaged graphene[S3] is probably a contributing factor, too. For hole doping ($\Delta V_g$ <-10 V), the contact resistance is much larger ($\approx$70 Ohm), yielding $Tr \approx 0.3$. This additional resistivity is due to reflection of charge carriers at pn junctions formed near the superconducting contacts.

## 3. Fabry-Pérot oscillations in the normal state

The term Fabry-Pérot (FP) interferometer refers to a cavity defined by two parallel semitransparent mirrors, in which monochromatic waves bouncing back and forth between the mirrors lead to interference and, therefore, resonances in transmission. The pronounced oscillations observed in resistance of our devices (Fig. 1b of the main text) are due to interference of electron waves partially reflected by the pn junctions formed near the nanostrip contacts[24,25].

For a pn junction with a smooth potential profile only incident waves almost perpendicular to the junction have a non-vanishing transmission probability[S4]. This determines the relative size, $\delta G$, of the peaks in $R_n$ which appear under the resonance condition, $2L^*/\lambda_F = N$ where $N$ is integer, corresponding to the formation of standing waves in a cavity of length $L^*$. Using the dispersion relation $\varepsilon_F = h v_F/\lambda_F$ ($\varepsilon_F$ and $v_F$ are the Fermi energy and velocity, respectively) the period of the standing-wave resonances on the energy scale $\varepsilon$ is expected to be $\varepsilon_0 = h v_F/2L^*$. Taking into account the energy-dependent contributions to the conductivity, $G(\varepsilon) = G_0 + \delta G \sin\frac{2\pi\varepsilon}{\varepsilon_0}$, the current $I$ flowing through the FP cavity is given by $I = \frac{1}{e}\int_{\varepsilon_F - eV_b/2}^{\varepsilon_F + eV_b/2} G(\varepsilon)d\varepsilon$, which yields oscillations in $I$ and the differential conductance $dI/dV_b$ in the form

$$I = G_0 V_b + \frac{\varepsilon_0}{\pi e}\delta G \sin\frac{2\pi\varepsilon_F}{\varepsilon_0}\sin\frac{\pi e V_b}{\varepsilon_0} \qquad \frac{dI}{dV_b} = G_0 - \delta G \sin\frac{2\pi\varepsilon_F}{\varepsilon_0}\cos\frac{\pi e V_b}{\varepsilon_0}. \qquad (S1).$$

The latter expression describes FB oscillations as a function of both $\varepsilon_F \propto n^{1/2} \propto |\Delta V_g|^{1/2}$ and $V_b$. Qualitatively, this is the behavior observed experimentally and shown in Fig. 1b of the main text.

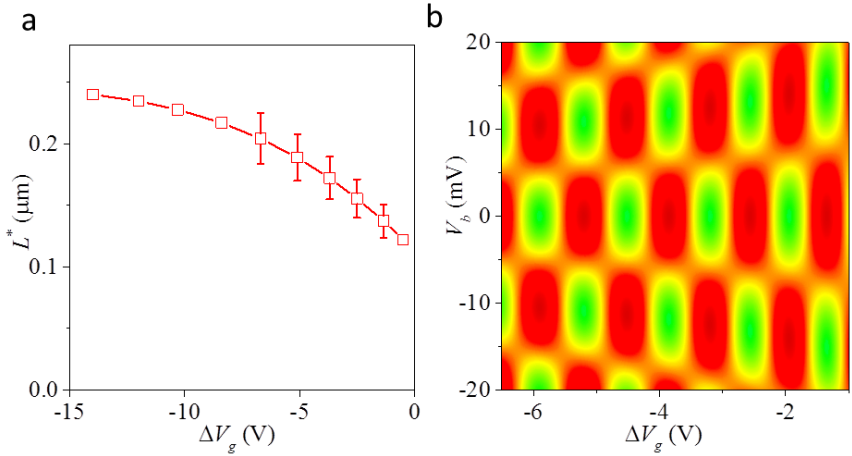

**Figure S3 | Changing length of graphene FP resonators.** (**a**) The effective length of the FP cavity as found from the periodicity of the resistance oscillations along the $V_b$ axis. The data points with error bars are from the plot shown in the inset of Fig. 1a of the main text. At low doping, the cavity is significantly shorter than the lithographically-defined distance between Nb contacts, $L \approx 0.25$ μm. (**b**) Modelling the chequered pattern found experimentally (Fig. 1b of the main text). In the calculations we used the cavity length found in (a) and a constant capacitance coupling to the back gate. The latter is $\approx 5 \times 10^{10}$ cm$^{-2}$ per V as determined experimentally from the period of Shubnikov-de Haas oscillations.

Despite the overall agreement, the experiment shows notable deviations from the exact periodicity expected in eq. (S1). They are not important in the context of this report but probably worth of pointing out. The observed deviations are due to changes in the effective position of pn junctions with varying graphene's doping. Indeed, one can see in the inset of Fig. 1b (also, Fig. S3b below) that the chequered pattern becomes stretched along the y-axis with approaching the NP. This indicates that $\varepsilon_0$ becomes progressively larger closer to the NP, which means that the effective length $L^*$ of our FP cavity becomes shorter with decreasing hole doping. Figure S3a plots the inferred values of $L^*$ for different $\Delta V_g$, which shows that the length of the FP cavity changes as much as by a factor of 2. This behavior is not unexpected. Indeed, graphene is electron doped by the contact with Nb/Ta and, as we increase $|\Delta V_g|$ and induce hole doping in graphene, the pn junctions are expected to become sharper and shift closer to the nanostrip contacts, approaching the limit $L^* = L$ at high doping. For completeness, we have also modelled changes in $R_n$ using eq. (S1) and the inferred changes in $L^*$. The results are plotted in Fig. S3b that shows good agreement with the detailed behavior observed experimentally in Fig. 1b of the main text.

## 4. Proximity superconductivity in the ballistic regime

In our SGS devices, the critical current increased with decreasing $L$, and even our shortest junctions with $L \approx 0.15$ μm were found not to be in the limit of short $L$, in which case $I_{c,e}$ should be independent of $L$. Figure S4 shows the observed supercurrent for different $L$ using data from 3 sets of devices, each having a different width and somewhat different $T_C$ and capacitance to the back gate. For clarity, we combine all the data on a single graph by plotting $I_c R_n/e\Delta$ because this product should be less sensitive to many experimental details, including $\Delta$ and doping (Fig. 2c of the main text). One can see that $I_c R_n$ varies approximately as $1/L$ and shows little evidence of saturation at short $L$.

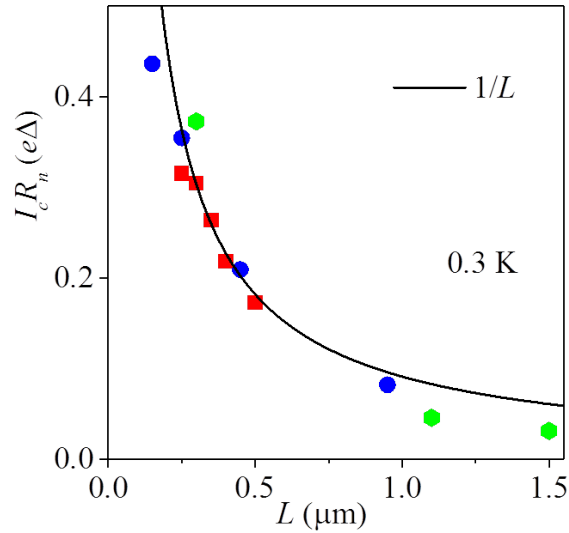

**Figure S4 | Effect of the junction length on supercurrent**. The data are for 12 ballistic devices with different $W$. Colors indicate sets of the devices with the same $W$: red 3 μm, blue 5 μm, green 6.5 μm. For each data set, $I_c(L)$ follows the same functional dependence as $I_c R_n$ because of $R_n$ was practically independent of $L$ for same $W$. For the two longest devices, the critical current falls below the extrapolated $1/L$ dependence probably because of external noise or fluctuations.

In the limit of long but diffusive junctions studied in literature[26], the gap $\Delta$ in eq. (1) of the main text should be substituted with the Thouless energy $E_T$ that varies as $1/L^2$. For our ballistic junctions, the square dependence cannot fit, even qualitatively, the experimental data in Fig. S4. We attribute the observed $1/L$ dependence to the energy scale analogous to $E_T$ but modified for ballistic transport. In this case, the broadening of Andreev bound states during their time, $\Lambda/v_F$, spent in graphene can roughly be estimated as $E_{T^*} = \hbar v_F/\Lambda$, where $\Lambda = L/Tr$. As discussed in the main text, for our devices, $E_{T^*}/\Delta \sim 1$. The absence of saturation in $I_{c,e} R_n$ at short $L$ indicates that $\Delta$ in eq. (1) should probably be substituted with $E_{T*}$, by analogy with long diffusive SNS junctions. Unfortunately, there is no theory for the case of ballistic Josephson junctions to support the above analysis and provide a numerical coefficient in front of $\hbar v_F/L$. Note that, for the case of graphene, such a ballistic theory should also take into account that transmission through pn junctions is strongly angle dependent because of Klein tunneling[S4,S5].

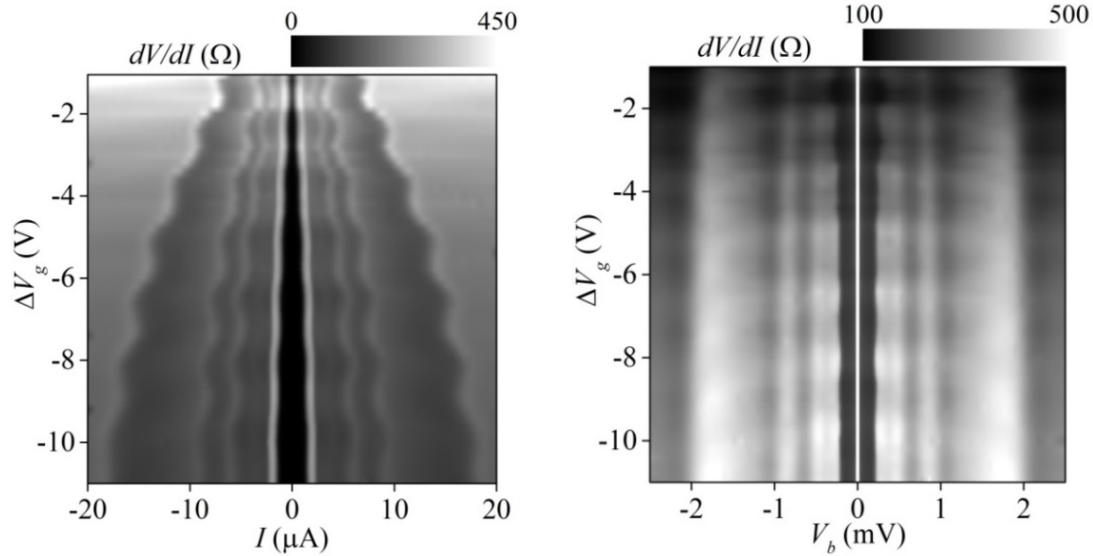

**Figure S5 | Sub-gap multiple Andreev reflection in a FP cavity.** (**a**) Differential resistance for hole doping of the device in Figs. 2a,b of the main text. Larger currents $I$ are used here. (**b**) The same characteristics as in (a) but plotted as a function of $V_b$ rather than $I$.

In principle, one can think of attributing the observed oscillations in $I_{c,e}$ to an oscillatory behavior of the gap $\Delta$ that enters the right part of eq. (1) of the main text. We found experimentally that this was not the case. Indeed, Fig. S5 shows how features in the differential I-V characteristics of our SGS devices evolve as a function of doping, current and bias. The FP transmission resonances result in a pronounced oscillatory pattern in Fig. S5a, which qualitatively mirrors oscillations in $I_c$ and can be attributed to multiple Andreev reflection (MAR)[1,28]. Figure S5b shows the same differential conductance but as a function of applied bias $V_b$. The peaks due to MAR occur at $V_b \approx 2\Delta/N$ (Fig. 2d of the main text). Their positions do not exhibit any discernable FP oscillations that would indicate oscillations in $\Delta$. The wavy pattern in Figure S5b is due to oscillatory broadening of MAR peaks (see Fig. 2d of the main text).

### 5. Reproducibility of proximity patterns in high magnetic fields

In the high-flux regime, the supercurrent randomly changed with varying $B$ and $\Delta V_g$. Fluctuating patterns such as the one shown in Fig. 3e of the main text were found to be stable over a period of several hours and reproducible if $B$ was swept up and down (Fig. S6). This proves that the observed fluctuations are not caused by flux creep in the adjacent superconducting contacts. Such creep can indeed appear due to movements of pinned vortices and is an irreversible process. Flux jumps could be observed over longer time scales and with varying $B$ over intervals larger than several mT. The flux instability is easily distinguishable leading to abrupt changes in proximity patterns as illustrated in Fig. S7.

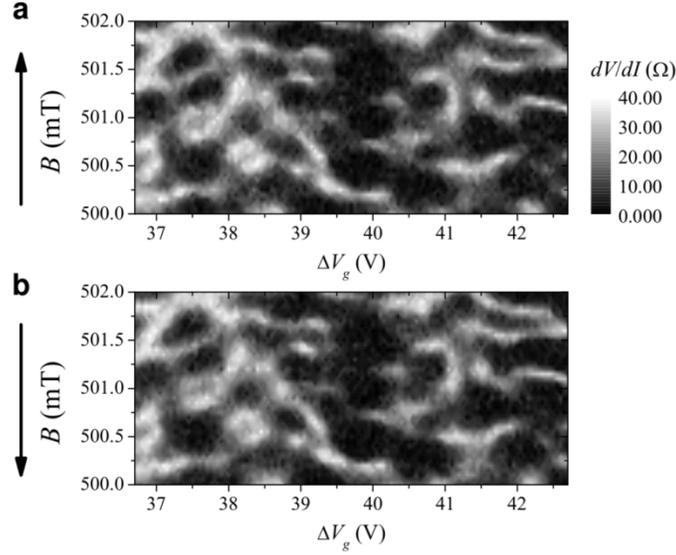

**Figure S6| Reproducibility.** Differential resistance maps measured by sweeping $V_g$ and gradually increasing (**a**) and decreasing (**b**) $B$ in steps of 0.1 mT. Time elapsed between the shown maps was 12 h.

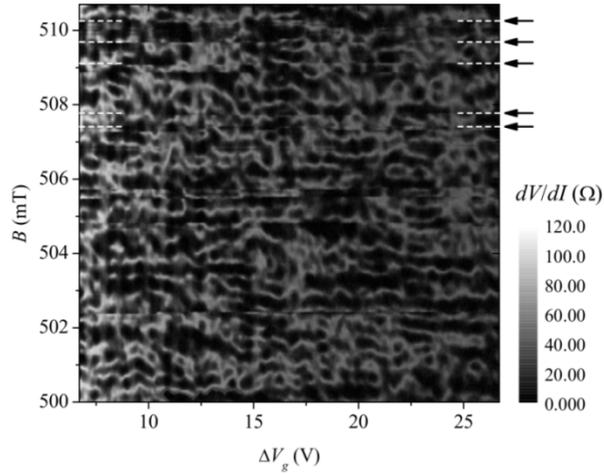

**Figure S7| Abrupt changes in a proximity pattern caused by flux creep.** The map was acquired over a period of 3.8 days. The arrows indicate some of the discontinuities caused by flux jumps between consecutive sweeps of $V_g$.

## 6. Correlation scales

Characteristic scales for the observed fluctuations in $I_c$ were calculated as follows. For a given set of $N$ resistance values $R_i$ measured at magnetic fields $B_i = B_0 + i\Delta B$ where $0 \leq i < N$ is an integer and $\Delta B$ is the spacing in $B$ between consecutive sweeps, the autocorrelation $K_n^{(B)}$ at a scale $\delta B = n\Delta B$ ($0 \leq n < N$ is an integer) is given by

$$K_n^{(B)} = \left[\sum_i \frac{(R_i)^2}{N}\right]^{-1} \sum_i \frac{R_i R_{i+n}}{N-n}.$$

Similarly, given a discrete set of $N$ resistances $R_j$ measured at Fermi energies $\varepsilon_j = \varepsilon_0 + j\Delta\varepsilon$ (where $j$ is an integer and $\Delta\varepsilon$ is the spacing of energies) the autocorrelation $K_n^{(\varepsilon)}$ for an energy scale $\delta\varepsilon = n\Delta\varepsilon$ is given by

$$K_n^{(\varepsilon)} = \left[\sum_j \frac{(R_j)^2}{N}\right]^{-1} \sum_j \frac{R_j R_{j+n}}{N-n}.$$

Figure 3e of the main text and Fig. S8a show two maps of the fluctuating proximity effect. For each of them, the found autocorrelations $K_n^{(B)}$ are averaged over all $\varepsilon_F$ to find $\langle K_n^{(B)} \rangle$, and the autocorrelations $K_n^{(\varepsilon)}$ are averaged over all $B$ to find $\langle K_n^{(\varepsilon)} \rangle$. The averaged autocorrelations are shown in Figs S8b,c. The plots yield a characteristic scale for suppression of $I_c$ with changing $B$ as ≈0.5 mT (Fig. S8b). It requires changes in $\varepsilon_F$ by ≈ 1.7 meV at low $n$ whereas somewhat smaller changes of ≈1 meV are required at high $n$ (Fig. S8c). The latter can be understood by longer $\Lambda$ at higher doping, which should lead to smaller $E_T^*$, as discussed in the main text.

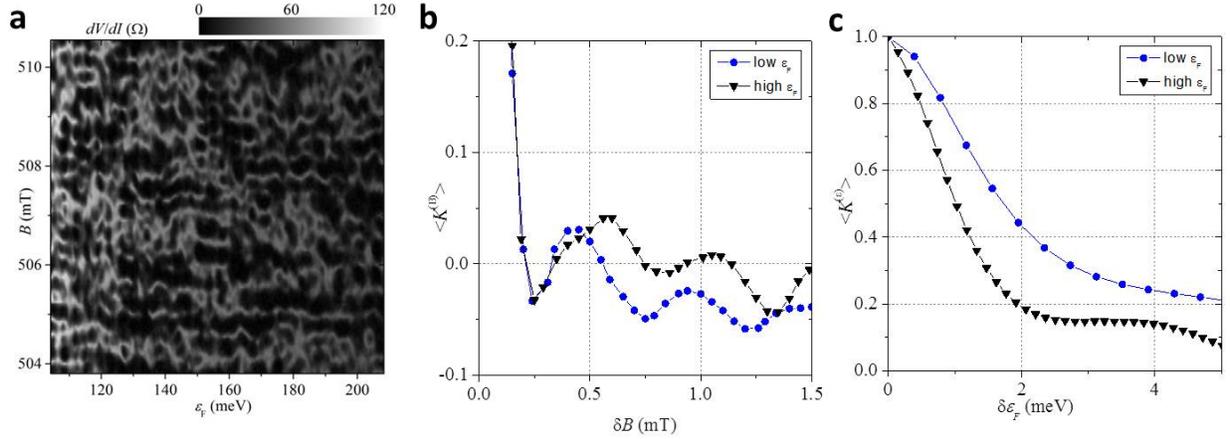

**Figure S8| Characteristic scales for high-field Andreev bound states.** (**a**) Another example of $dV/dI$ maps, covering a different range of doping with respect to the map in Fig. 3e of the main text. (**b**) and (**c**) are autocorrelation functions $\langle K^{(B)} \rangle$ and $\langle K^{(\varepsilon)} \rangle$, respectively, for maps in (a) and of Fig. 3e, which are labelled as low and high $\varepsilon_F$, respectively. The large peak at zero $\delta B$ arises due to a finite interval (0.1 mT) between consecutive sweeps.

### 7. Andreev bound states in zero and finite magnetic fields

The supercurrent through a normal metal placed between two superconductors is mediated by a process known as Andreev reflection[27]. In this process (Fig. S9a), an electron arriving at the NS interface forms a Cooper pair with another electron found near the Fermi energy $\varepsilon_F$, and this sends a hole back into the Fermi sea of the normal metal. The transfer of a Cooper pair through an SNS junction requires two such reflections at the opposite NS interfaces ('double' Andreev process, in which the involved electrons (*e*) and holes (*h*) retrace each other's trajectories). This leads to the formation of so called Andreev bound states.

If *e* and *h* forming Andreev states have exactly opposite momenta (***p*** = -***p'***), their phases acquired along trajectories inside the normal metal cancel exactly. Andreev bound states can also be formed by *e* and *h* with slightly different momenta, provided the carriers reside within the superconducting gap $\Delta$ (that is, $|p - p'|v_F < \Delta$) and the phase shift acquired along the retracing *e-h* trajectories, $\delta = |p - p'|\Lambda/\hbar$, is small. This leads to the known constraint, $\Delta < \hbar v_F/\Lambda = E_T^*$. Andreev-type trajectories with longer effective length $\Lambda$ do not contribute to the transfer of Cooper pairs.

Following a consideration similar to the above, one can find that *e* and *h* involved in the formation of Andreev bound states do not have to retrace each other exactly and may have slightly misaligned trajectories as illustrated in the top part of Fig. S9a. The conversion of two electrons from a 2D metal into a Cooper pair necessitates the condition $p_y = -p'_y$ (indices $x$ and $y$ refer to the directions perpendicular and parallel to the GS interface, respectively). On the other hand, restrictions on the $x$-components of the momenta [$p_x = (\varepsilon/v)\cos\theta$ and $p'_x = (\varepsilon'/v)\cos\theta'$] are set by the requirement $\left|v_F\sqrt{p_x^2 + p_y^2} - \varepsilon_F\right| < \Delta$, which means that energies of the charge carriers involved in Andreev bound states should reside within the gap. This sets the following constraint,

$$\cot\theta' - \cot\theta \approx \frac{\theta'-\theta}{\sin^2\theta} < \frac{\Delta/(v_F\cos\theta)}{p\sin\theta},$$

on the misalignment angle, $\delta\theta = \theta - \theta'$, between ballistic $e$ and $h$ trajectories forming Andreev bound states (Fig. S9a). The above expression can be simplified as

$$\delta\theta < \frac{\Delta}{\varepsilon_F}\tan\theta \qquad (S2).$$

Another important requirement is that the ends of $e$-$h$ trajectories should not be farther away from each other than max{$\xi$, $\lambda_F$} (see Fig. 9a). Otherwise, two electrons cannot form a Cooper pair inside a superconductor, where its size is given by the correlation length $\xi$. On the other hand, positions of two electrons within a normal metal are indistinguishable if they are separated by less than $\lambda_F$. For all carrier densities in our experiments, $\xi < \lambda_F$, which results in the following condition, $\frac{L\delta\theta}{\cos^2\theta} < \lambda_F$. Finally, one more limitation is set by the requirement that the phase shift between an electron and Andreev-reflected hole should be small, which leads to $(\sec\theta' - \sec\theta)L/\lambda_F \approx \frac{L\delta\theta\sin\theta}{\lambda_F\cos^2\theta} < 1$. The latter two constraints are nearly identical and require

$$\delta\theta < \frac{\lambda_F}{L}\cos^2\theta. \qquad (S3)$$

Under our experimental conditions $\Delta \approx E_{T*} = \hbar v_F/L$ (see the main text), constraints (S2) and (S3) have similar strengths. Note that eq. (S2) discriminates against Cooper pairs transferred perpendicular to the NS interface, whereas eq. (S3) against those at shallow angles. For our devices, we estimate that Andreev bound states with $\theta$ ~1 and $\delta\theta$ less than a couple of degrees should dominate Cooper-pair transport.

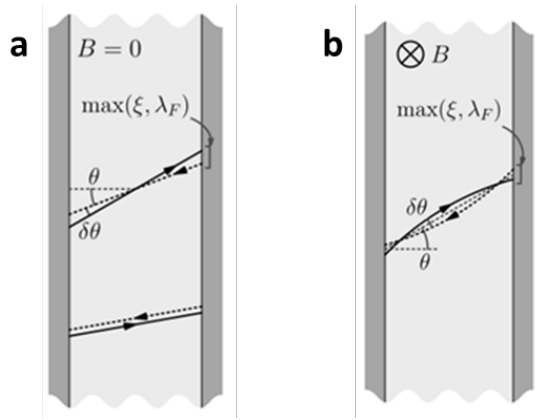

**Figure S9| Andreev bound states and allowed misalignment of contributing trajectories. (a)** The bottom set of *e-h* trajectories illustrates the standard double Andreev process. Top: Slightly misaligned *e-h* trajectories can also

form Andreev bound states if their positions at the two NS interfaces are spatially close. The constraints are given by equations (S2) and (S3). (**b**) Similarly, slightly curved cyclotron trajectories ($r_c \gg L$) can form Andreev bound states with constraints set by eq. (S4).

The supercurrent provided by Andreev bound states is suppressed by magnetic field. In low *B*, the dominant effect is interference between Cooper pairs that cross the normal metal along different paths. If $BS \sim \phi_0$, Cooper pairs acquire broadly distributed phase shifts, and this leads to the oscillatory suppression of the supercurrent as described by eq. (3) of the main text.

In ballistic devices (that is, with large *L*), magnetic field curves *e-h* trajectories, leading to their misalignment such that Andreev-reflected electrons and holes can no longer retrace each other exactly[22,S6]. The effect is rather similar to the zero-*B* misalignment described above but is caused by a finite cyclotron radius, $r_c = p_F/eB = \hbar\sqrt{\pi n}/eB$ (Fig. S9b). For *e-h* trajectories leaving a superconducting contact at an angle θ (Fig. S9b), the cyclotron curvature leads to misalignment

$$\delta\theta \sim \frac{L}{r_c \cos\theta} \qquad (S4).$$

Combined with the constraint set by (S2), eq. (S4) yields $r_c > \frac{\varepsilon_F}{\Delta} L$, in order to support Andreev bound states at finite *B*. This condition is satisfied if $B < B^*$ where

$$B^* \sim \frac{\Delta}{eLv_F} \qquad (S5).$$

For our ballistic SGS devices with submicron *L*, $B^*$ is a few mT. For $B > B^*$, it becomes impossible for Andreev-reflected electrons and holes to form closed loops that are necessary to transfer Cooper pairs. Accordingly, the supercurrent in the graphene bulk is suppressed.

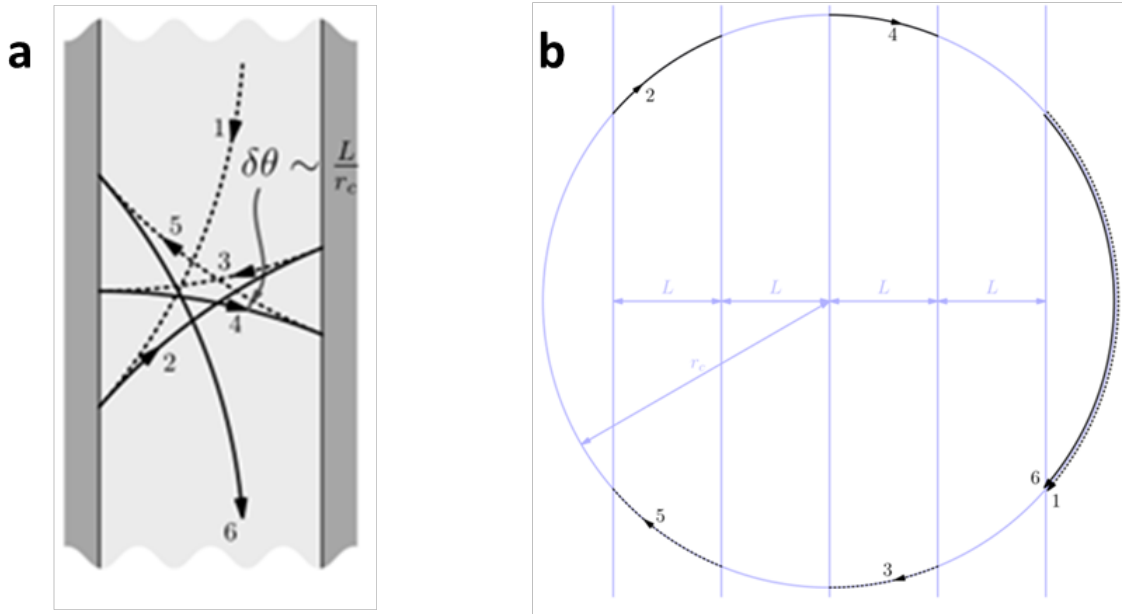

**Figure S10| No Andreev bound states for strongly curved trajectories.** (**a**) Typical trajectory formed by Andreev-reflected electrons and holes for $B > B^*$. (**b**) Ballistic orbits in (a) correspond to well-defined sectors of the full cyclotron orbit. Labelling of the segments is the same in both plots.

One may wonder whether Andreev bound states can be formed for $B \gg B^*$ by a fortuitous combination of a number of segments of cyclotron orbits as charge carriers bounce between two superconducting contacts. The answer is No. As shown in Fig. S10, the condition $B \gg B^*$ results in long open trajectories. Indeed, each Andreev reflection process involves two segments of the full cyclotron orbit, one for an electron and the other for a hole (Fig. S10b). Each consecutive reflection increases the deflection angle by $\delta\theta \sim L/r_c$, until the final segment (#1 and 6 in Fig. S10) brings the quasiparticle back to the same NS interface where the last Andreev reflection took place. After this step, a reversed sequence of Andreev reflections follows, transferring the charge in the opposite direction. This results in an infinite path containing periodic sets of star-shaped $e$-$h$ orbits bouncing between the superconducting leads. As shown in Fig. S10b, the number $N$ of Andreev processes linking sets of ($N$ +1)-pointed stars is determined simply by int{$2r_c/L$}. The shape of such open trajectories is quite generic: they describe the electron drift along a graphene strip. It is also worth of mentioning that, depending on partition of the cyclotron orbit in Fig. S10b, such drifts can be in both 'up' and 'down' directions in Fig. S10a. Among such star-shaped orbits, there is a special one that has a zero drift velocity and, hence, it is closed. Nonetheless, even the special orbit cannot support an Andreev bound state, because half such an orbit provides the electron (hole) transfer from one S contact to the other whereas the other half brings it back.

Although closed Andreev trajectories are forbidden in the ballistic bulk for $B \gg B^*$, they are still allowed near graphene edges. Two examples of such orbits are shown in Fig. S11, and many others can be drawn depending on scattering details and $B$. These near-edge orbits have closed ends at both NS interfaces. This means that, despite different lengths of $e$ and $h$ parts (solid and dashed curves in Fig. S11), the $e$-$h$ trajectories transfer Cooper pairs between the superconducting contacts. For each Andreev state, its current carrying capacity is given by eq. (2) of the main text.

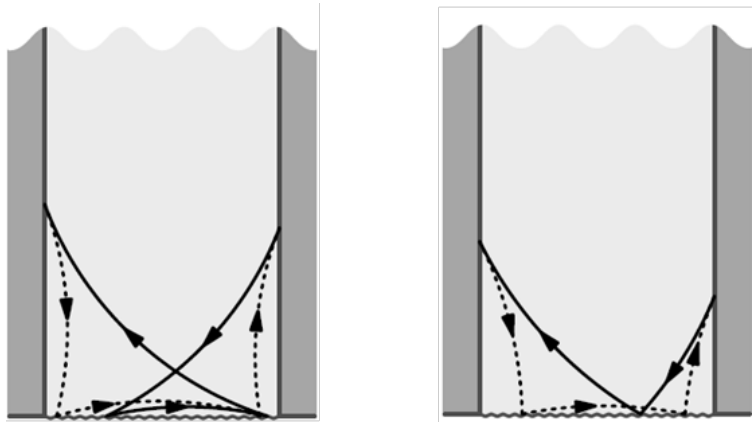

**Figure S11| Andreev bound states near graphene edges.** Examples of closed Andreev trajectories due to diffusive scattering at graphene edges.

Finally, the situation changes again if $r_c < L/2$, that is, the cyclotron orbit becomes small enough to fit within the graphene strip without touching the NS interfaces. In this case, which takes place in $B > \frac{2\hbar}{eL}\sqrt{\pi n(V_g)}$, charge carriers can be transferred between the contacts only by skipping orbits and only in one direction at each of the two graphene edges (Fig. 3i of the main text). As a result, the transfer of Cooper pairs is no longer possible anywhere, either along graphene edges or in the bulk.

## 8. High field cutoff in the proximity effect

It was argued above and indicated in the main text (Fig. 3d) that random Andreev bound states could survive in high $B$ until cyclotron orbits start fit between superconducting contacts. To further substantiate this experimentally, Figure S12 shows the differential resistance $R$ measured over a large range of $B$ and $V_g$. Three different regions can clearly be distinguished for the case of positive $\Delta V_g$ (electron doping; no pn junctions at the Nb contacts). One of the regions corresponds to the conventional Josephson effect and is found in a narrow interval of small $B$ (black stripe in Fig. S12a). Here the cyclotron radius $r_c \gg L$, and the proximity is mediated by practically straight Andreev bound states, that is, $B < B^*$ (Fig. S12b). In high $B$, our ballistic devices enter the opposite regime, $r_c \ll L$ (Fig. S12d), which results in skipping trajectories and Shubnikov-de Haas oscillations, characteristic of the quantum Hall regime in the two-terminal geometry. In between the two extremes lies a wide range of $B$ and $\Delta V_g$ in which pockets of the proximity superconductivity were observed (blue region). Boundaries between the three regimes are clearly seen due to changes in color in Fig. S12a. From the high-$B$ side, the boundary is well described by the condition $2r_c = L$ which is shown by the black curve. In the blue region, the proximity effect randomly occurs all the way up to the high-$B$ boundary (see Fig. 3d of the main text). A finite resistance that appears in the blue region of Fig. S12a is due to sampling and averaging over relatively large intervals of $B$. On this scale (>> 1mT), individual superconducting states such as in Figs. 3e and S8a cannot be resolved but their occurrence frequency is reflected in different shades of blue.

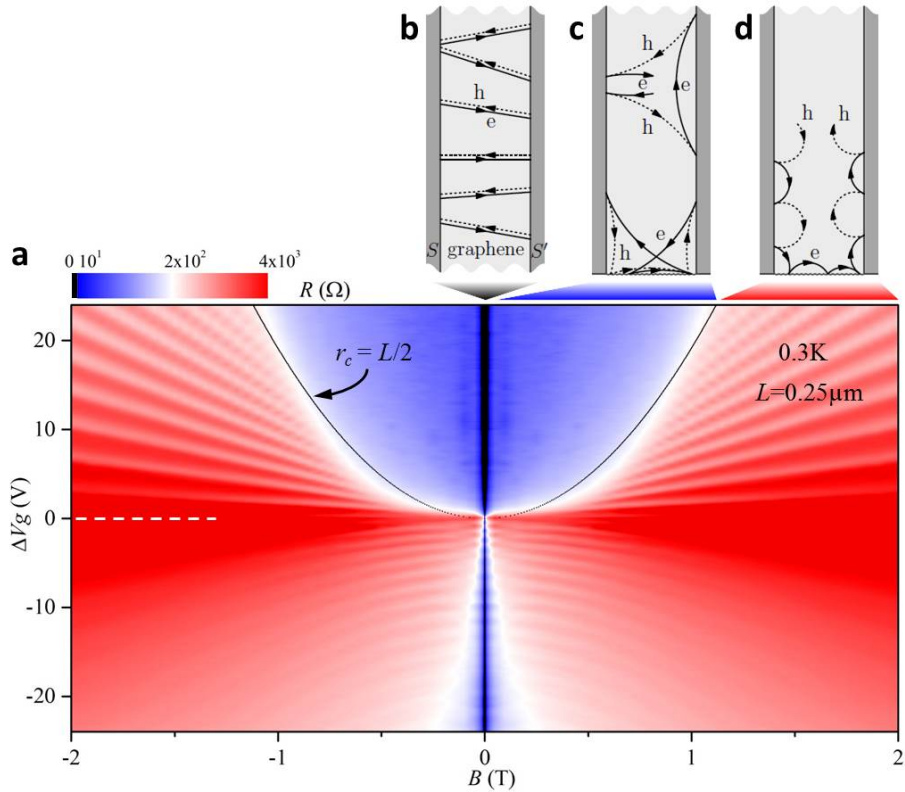

**Figure S12| Characteristic regimes in ballistic Josephson junctions.** (**a**) Resistance map with the probing current of 2 nA. The Josephson effect mediated by the conventional Andreev bound states shows up in black. Blue: Pockets of proximity superconductivity are observed for electron doping. (**b-d**) Sketches of electron and Andreev-reflected hole trajectories in graphene in low (b), intermediate (c) and high $B$ (d). In (b), graphene can support large supercurrents. In (c), the supercurrent is supressed because the cyclotron motion prevents $e$ and $h$

trajectories from retracing each other. In this case, the proximity can still be mediated by Cooper pairs that cross the junction near graphene edges. In the high-*B* regime (d), no Andreev states are possible.

Note that high-*B* pockets of the superconducting proximity could not be observed for hole doping. Instead clear quantum oscillations cover nearly the entire interval of *B* and *n* (Fig. S12a). The latter oscillations continue first as an extension of FP oscillations in low *B* (see Fig. 2b of the main text) but then they exhibit a phase shift and start bending. This behavior is attributed[24,25] to Klein tunneling through the hole-doped graphene strip between n-doped contact regions. It is important to note that Klein tunneling collimates electron and hole trajectories perpendicular to pn interfaces[S4,S5]. To form near-edge Andreev states shown in Fig. S11, it requires cyclotron trajectories tilted towards the NS interface and, therefore, the Klein-tunneling collimation is expected to strongly suppress such Andreev states. This is likely to be the reason that no high-*B* proximity states could be observed in this regime.

## 9. Spatial distribution of supercurrents

It has been shown in literature that zigzag segments at graphene edges may lead to enhanced conductance, even for strongly disordered, non-crystallographic egdes[S7,S8]. Furthermore, in the strip geometry, an extended back gate can causes inhomogeneous doping for purely electrostatic reasons[S9]. Therefore, it is reasonable to ask whether the discussed near-edge Andreev states can also be enhanced by such mechanisms that influence near-edge conductance in the normal state. We find no evidence for this. Our results indicate that the supercurrent was uniformly distributed across the entire width of our SGS junctions. Indeed, the Fraunhofer pattern such as in Fig. 3a of the main text is given by the Fourier transform of a spatial distribution of the supercurrent across an SNS junction[1,S10]. For the reported devices, all the Fraunhofer patterns could be fit well, assuming a uniform supercurrent distribution (Figs. 3a and S13). If the proximity effect in low *B* were enhanced near graphene edges by one of the mentioned mechanisms, one should expect the corresponding signature in the Fraunhofer pattern. For weak inhomogeneity, this would result in a narrowing of the central peak, and it would split into two if the near-edge supercurrent becomes dominant[S10]. As shown in Fig. S13, no narrowing of the central Fraunhofer peak could be detected at any doping including the NP.

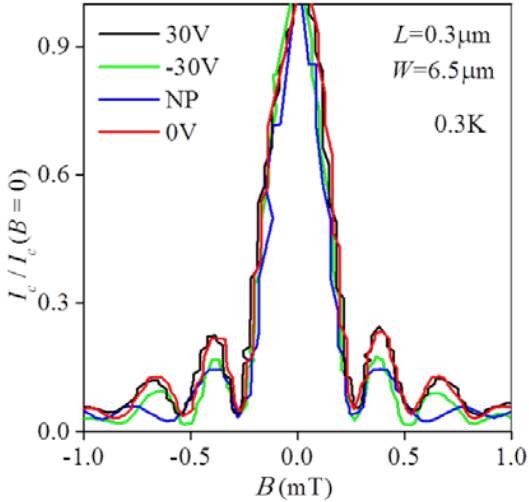

**Figure S13| Uniform supercurrent distribution in low magnetic fields.** Fraunhofer patterns in our SGS junctions for several representative gate voltages. The central peak always has a width twice the others, which rules out any significant contribution from supercurrents flowing near graphene edges for the case of $B < B^*$.